# SIREN Cytoscape plugin: Interaction Type Discrimination in Gene Regulatory Networks

**Jason Montojo[1], Pegah Khosravi[2,1], Vahid H. Gazestani[3], Gary D. Bader[1]***

[1]The Donnelly Centre, University of Toronto, Toronto, Canada
[2]School of Biological Science, Institute for Research in Fundamental Sciences (IPM), Tehran, Iran
[3]Institute of Parasitology, McGill University, Montreal, Quebec, Canada

* Corresponding authors: Gary D. Bader – E-mail: gary.bader@utoronto.ca

**Abstract**
Integrating expression data with gene interactions in a network is essential for understanding the functional organization of the cells. Consequently, knowledge of interaction types in biological networks is important for data interpretation. Signing of Regulatory Networks (SIREN) plugin for Cytoscape is an open-source Java tool for discrimination of interaction type (activatory or inhibitory) in gene regulatory networks.

Utilizing an information theory based concept, SIREN seeks to identify the interaction type of pairs of genes by examining their corresponding gene expression profiles. We introduce SIREN, a fast and memory efficient tool with low computational complexity, that allows the user to easily consider it as a complementary approach for many network reconstruction methods.

SIREN allows biologists to use independent expression data to predict interaction types for known gene regulatory networks where reconstruction methods do not provide any information about the nature of their interaction types. The SIREN Cytoscape plugin is implemented in Java and is freely available at http://baderlab.org/Software/SIRENplugin and via the Cytoscape app manager.

# Findings
## Introduction

Integrating expression data with gene interactions in a network is essential for understanding cellular process. Many tools within the Cytoscape ecosystem were developed to analyse and interpret biological networks [1]. The SIREN Cytoscape app is a standalone tool for making fast and efficient gene interaction type determination. The app implements the SIREN algorithm which uses association-based approach to derive interaction type from expression data.

It has been shown that signed gene co-expression networks provide a sound foundation for module-centric analysis, network simulation and is essential for data interpretation [2, 3]. Analysis of gene expression data and other omics data commonly use co-expression methods. Most co-expression measures are categorized as either correlation coefficient or mutual information (MI) measures. MI-based methods, in particular, are used to measure non-linear associations and able to detect additional correlations invisible to the linear Pearson coefficient [4-6].

SIREN uses an information-based approach. This new method is capable of identifying the type of interaction between two interacting genes. It is important to note SIREN detects the interaction types but it cannot detect the direction of interactions. The resulting signed network of relationships among genes is available as a signed Cytoscape network for further analysis (Figure 1).

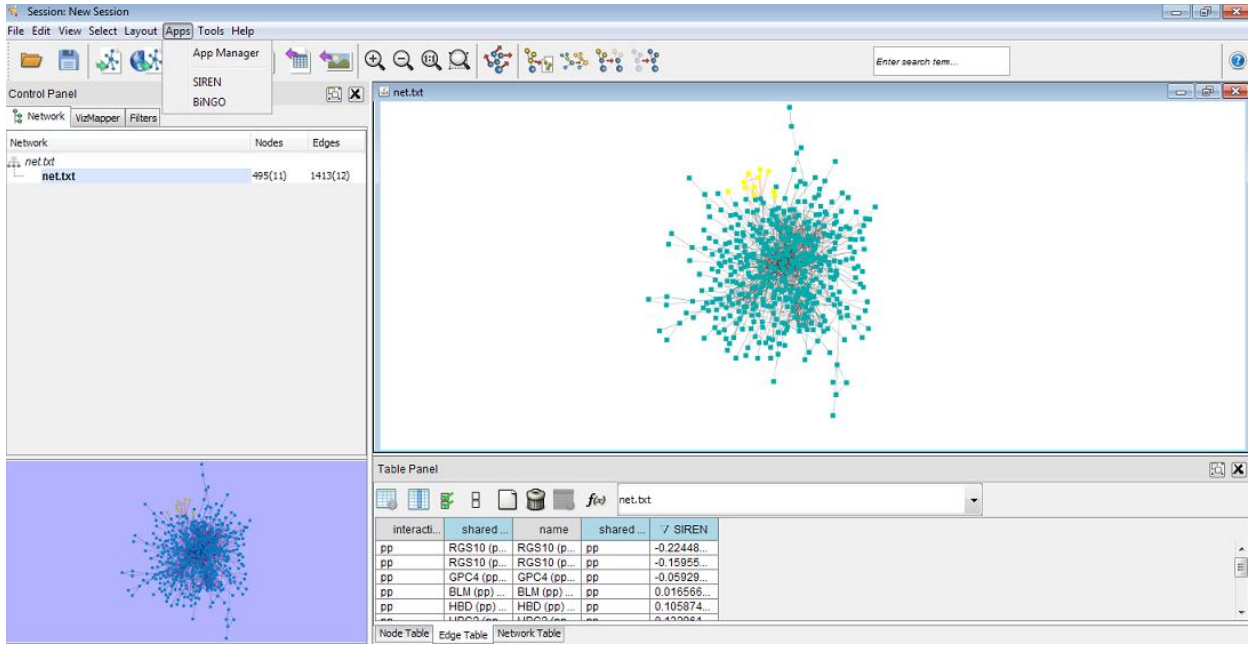

**Figure 1: The resulting signed network**
SIREN applied on reconstructed prostate cancer network. Network contains 495 nodes and 1413 edges, laid out in 1s.

**Methods and implementation**
The fundamental idea in SIREN is that if two connected nodes in a network have similar expression patterns, there is probably synergy between them. On the other hand, if their expression patterns are distinct, the interacting genes probably have a negative influence on each other [7]. Our method can be useful for the analysis of reconstructed networks and is unaffected by the choice of reconstruction method used to generate the input network. As inputs, SIREN uses expression data and a genetic network. SIREN determines a regulation type for every pair of connected nodes in the network by comparing the expression profiles of each pair and assigning a similarity score to it. To determine the distance between two nodes, SIREN rescales the raw similarity according to a rescaling matrix, and then integrates this score with a derivative form of point-wise mutual information formula.

$$SIREN = \sum W(x,y)p(x,y)\log(p(x,y)/p(x)p(y))$$

Where $W(x,y)$ is the rescaling matrix, $p(x,y)$ is the joint probability of a particular co-occurrence of events and $p(x)p(y)$ is the marginal probability. Using a reliable cut-off threshold (±0.158) which SIREN does not detect any interaction type in random data to evaluate this SIREN score, a regulation type (positive or negative) is assigned to the edges in the network. For more information please read the original paper in [7].

We implemented SIREN algorithm as an app for Cytoscape, which is a collection of powerful tools for network analysis and inference. It can be installed using Cytoscape's menu-driven app manager. Upon first use, the user must download the latest version of SIREN. The user then imports the network from file and uses the SIREN app to score the network. Finally, the software asks the user to import expression data from file. The network should be in two column format. There is no limitation for the number of row and column of expression data but the first column should contains IDs related to each gene which is the same with IDs in network [Additional file 1 and Additional file 2). SIREN does not impose any limitations for gene identifier but it should be the same for both the expression data and network. The plugin can determine interaction type for all nodes with available expression data. The resulting signed network can also be exported in standard formats, e.g. XGMML, SIF and PDF.

Users can update networks with their favorite threshold. Cytoscape's customizable filters of allow them to select edges with the highest SIREN score among all edges in a network. Users can easily add their own attributes to the networks with different colors based on activatory and inhibitory activity or SIREN weight. For example, the user can use a red line for activatory or blue line for inhibitory and different thickness of edges based on SIREN score for each edge (Figure 2).

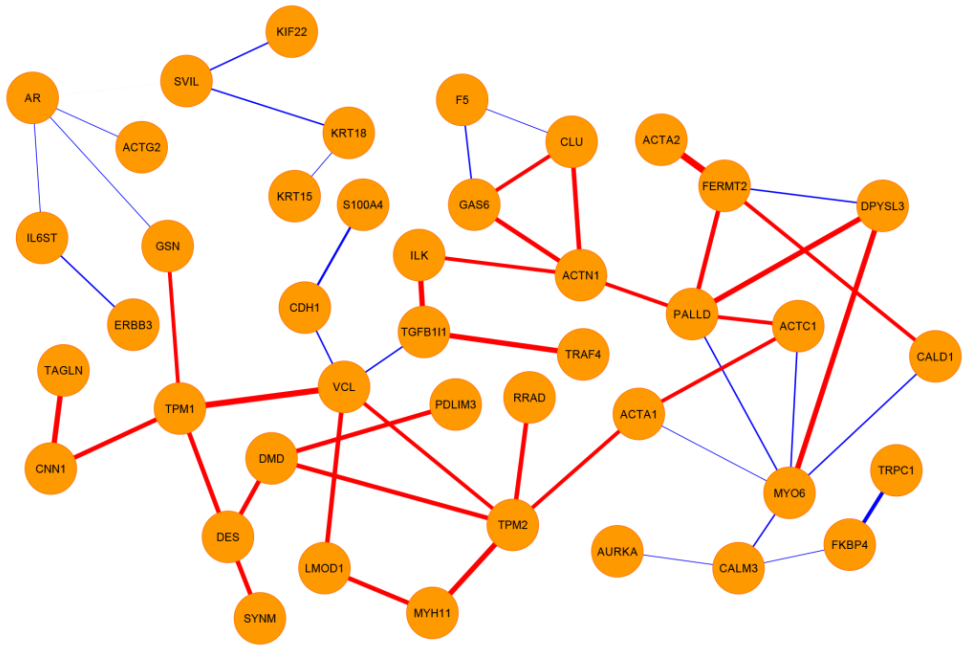

**Figure 2: Add attributes to the networks**
The blue edge shows inhibitory and the red one shows activatory. Also, the thicker edge width illustrates higher SIREN score and vice versa.

The size of the network data is limited by the amount of available memory and disk space. We recommend using a system with at least 4GB of total RAM.

To evaluate time and memory efficiency of SIREN, we used human prostate cancer network contains 1436 interactions of 526 genes from STRING platform [8] and microarray data were extracted from GEO (GDS2545), we tested SIREN on the derived network. We also evaluated SIREN efficiency on *E. coli* data contains 4005 interactions among 1696 genes that the network extracted from the RegulonDB database [9] and microarray dataset consisting of 907 samples from the Many Microbe Microarray Database ($M^{3D}$) Web site [10]. Detecting interaction types during these data took just 1 second on an Intel Core i5 system with 4GB of RAM. These examples show how the SIREN Cytoscape plugin can be used to determine interaction types for genes that may be involved in prokaryotic and eukaryotic gene regulatory networks (GRNs).

## Conclusions

We described a statistical framework algorithm for inferring the regulatory type of interactions in a known GRN given corresponding genome-wide gene expression data. The algorithm is highly flexible and can be adopted by many reconstructed gene regulatory network with a number of large-scale data. For example, user can easily determine the interaction association (positive or negative) by integrating their reconstructed networks with available data from different sources.

## Availability and Requirements

Project name: SIREN app
Project home page: http://baderlab.org/Software/SIRENplugin
Operating system: Platform independent
Programming language: Java
Other requirements: Cytoscape version 3.0.0 or newer, Java SE 7
License: GNU LGPL
Any restrictions to use by non-academics: None

## Acknowledgements

This work was supported by NRNB (U.S. National Institutes of Health, National Center for Research Resources grant number P41 GM103504). PK is supported by the School of Biological Sciences of Institute for Research in Fundamental Sciences (IPM). VHG is supported by CIHR Systems Biology Fellowship. The authors would like to warmly thank the SIREN team for support this project.

## Author details

[1] The Donnelly Centre, University of Toronto, Toronto, Canada. [2] School of Biological Sciences, Institute for Research in Fundamental Sciences (IPM), Tehran, Iran. [3] Institute of Parasitology, McGill University, Montreal, QC, Canada.

## List of abbreviations

GRN: Gene regulatory network; MI: Mutual information; $M^{3D}$: Many Microbe Microarray Database; SIREN: Signing of Regulatory Networks

## Authors' contributions

JM and VHG developed the code for SIREN. PK analyzed the data and created the figures. PK and JM wrote the paper. GDB headed the research program. All authors made substantial written contributions to the manuscript, and have given approval to the final version presented here.

## Competing interests

The authors declare that they have no competing interests.

High resolution versions of all figures, associated details and Supplementary Information available at http://baderlab.org/PegahKhosravi/SIREN